%% file: stat.tex
\documentclass[useAMS,usenatbib,a4paper]{mn2e}

\pdfoutput=1

\usepackage{aas_macros}

\usepackage{graphicx}
\usepackage{mathptmx}
\usepackage{amssymb}

\newcommand{\<}{\begin{eqnarray}}
\renewcommand{\>}{\end{eqnarray}} 
\renewcommand{\bar}{\overline}
\renewcommand{\tilde}{\widetilde}
\renewcommand{\hat}{\widehat}
\newcommand{\Msun}{\rmn{M}_\odot}
\newcommand{\MAX}{\mathrm{MAX}}
\newcommand{\MIN}{\mathrm{MIN}}
\newcommand{\crit}{\mathrm{crit}}
\newcommand{\MML}{\mathrm{MML}}

\title[Estimators and goodness-of-fit tests for power laws]{
Estimators for the exponent and upper limit, and goodness-of-fit tests for (truncated) power-law distributions
}
\author[Th. Maschberger and P. Kroupa]
{Thomas Maschberger$^{1,2}$ and Pavel Kroupa$^1$\thanks{e-mail: tmasch@astro.uni-bonn.de, pavel@astro.uni-bonn.de\newline	
A computer program for data analysis is available from http://www.astro.uni-bonn.de/downloads }\\
$^1$ Argelander-Institut f\"ur Astronomie, Auf dem H\"ugel 71, Bonn, Germany\\
Rhine Stellar Dynamics Network (RSDN),\\
$^2$ Institute of Astronomy, Madingley Road, Cambridge CB3 0HA, United Kingdom}
\date{MNRAS accepted}

\pagerange{\pageref{firstpage}--\pageref{lastpage}} \pubyear{2008}

\begin{document}

\maketitle

\label{firstpage}

\begin{abstract}
Many objects studied in astronomy follow a power law distribution function, for example the masses of stars or star clusters.
A still used method by which such data is analysed is to generate a histogram and fit a straight line to it.
The parameters obtained in this way can be severely biased, and the properties of the underlying distribution function, such as its shape or a possible upper limit, are difficult to extract.
In this work we review techniques available in the literature and present newly developed (effectively) bias-free estimators for the exponent and the upper limit.
Furthermore we discuss various graphical representations of the data and powerful goodness-of-fit tests to assess the validity of a power law for describing the distribution of data.
As an example, we apply the presented methods to the data set of massive stars in R136 and the young star clusters in the Large Magellanic Cloud.
For R136 we confirm the result of \citet{koen2006} of a truncated power law with a bias-free estimate for the exponent of $2.20 \pm 0.78$ / $2.87 \pm 0.98$ (where the Salpeter-Massey value is $2.35$) and for the upper limit of $143\pm9 \Msun$ / $163\pm 9 \Msun$, depending on the stellar models used.
The star clusters in the Large Magellanic Cloud (with ages up to $10^{7.5}$ yr) follow a truncated power law distribution with exponent $1.62\pm0.06$ and upper limit $68\pm12 \times 10^3 \Msun$.
Using the graphical data representation, a significant change in the form of the mass function below $10^{2.5} \Msun$ can be detected, which is likely caused by incompleteness in the data.

\end{abstract}

\begin{keywords}
methods: statistical -- methods: data analysis -- stars: luminosity function, mass function -- galaxies: star clusters
\end{keywords}

\section{Introduction}
Many astronomical objects are distributed according to a power law.
The probably most prominent example is the mass function of stars more massive than 0.5 $\Msun$ with the Salpeter-Massey exponent of 2.35.
Further examples are the mass functions of young star clusters and of molecular clouds.
Modern observational techniques and state-of-the-art models provide data such as stellar masses with unprecedented accuracy.
However, the statistical analysis of those data is not yet always optimal.
The technique of binning the data suffers from losing a lot of information.
The grouping of data into cells instead of using every data point obscures details of the observed distribution.
This is an especially serious problem in the upper range, where the bins are only sparsely filled.
Furthermore the obtained estimates of the slope can be severely biased \citep[see e.g.][]{maizapellaniz+ubeda2005}.
A method based on a  particular graphical display of the data which avoids grouping and allows one an estimate of the upper limit was given by \citet{koen2006}.
Another successful approach is to use the Maximum Likelihood method, which has been applied by \citet{jauncey1967} on extragalactic radio sources.
\citet{crawford-etal1970} derived a Maximum Likelihood estimator for the exponent without grouping the data and including an upper limit.

A further step in data analysis, equally important as estimating the parameters, is the validation of the assumed power law form of the distribution.
The simplest way to do this is to look at the histogram of the data in a double logarithmic plot.
If this plot appears to be linear then the consistency of the data with a power law is concluded.
But the significance of deviations from linearity are hard to state in an objective way by mere visual inspection.
A further, more elaborate way is to apply a goodness-of-fit test such as the Kolmogorov-Smirnov test.
If the calculated test statistic lies in some acceptance range then also consistency is concluded.
But it is possible that the test statistic calculated with data stemming from an alternative hypothesised distribution similar to the power law might as well fall in the acceptance range.
The test then fails to produce the right result since it has not enough ``power'' to discriminate.
Therefore the ``power'' properties of a goodness-of-fit test have likewise to be examined.
Such a study is -- to our knowledge -- not yet available in the astronomical literature.

In this work we describe estimation methods and compare their biases and variances (Section \ref{estimators}).
Since the data may stem from a truncated power law we focus on estimators which can be used in this case.
In the second part (Section \ref{secgoodness}) we investigate the question whether the data are consistent with the assumed power law distribution.
As informal aids to answer this we discuss various plotting recipes (Section \ref{plottingmethods}).
For an objective decision we present goodness-of-fit tests with a study of their discrimination power (statistical power) under the hypotheses of a truncated and infinite power law.
Finally, in Section \ref{examples}, we will apply the introduced methods on the massive stars in R136 and the young star clusters in the Large Magellanic Cloud.

\section{General results, definitions and notation}

\subsection{The power law distribution}
\label{powerlaw}
In this work we only consider  power law distribution functions (DF) with a negative exponent, $-\alpha$ ($\alpha > 1$).
By convention the sign is separated from the absolute value.
Besides the exponent such a distribution is further parametrised by the lower and upper limit, $x_\MIN$ and $x_\MAX$.
The probability density is then given by
\< p (x; \alpha, x_\MIN, x_\MAX) &=& \frac{ 1 - \alpha }{ x_\MAX^{1-\alpha} - x_\MIN^{1-\alpha} }  x^{-\alpha} \label{pltruncpdf}, \>
and the cumulative distribution function (DF) is
\< P (x) &=& \frac{ x^{1-\alpha} - x_\MIN^{1-\alpha} }{ x_\MAX^{1-\alpha} - x_\MIN^{1-\alpha} }. \label{pltrunccdf} \>

The family of distributions given by eq. \ref{pltruncpdf} includes the ``infinite'' or ``not truncated'' distributions with infinite upper limit, $p(x;x_\MAX=\infty) := p_\infty (x)$, which is also known in the (non-astronomical) literature as the Pareto distribution.
The density function reads then
\< p_\infty (x) &=& -\frac{1- \alpha}{x_\MIN^{1-\alpha}} x^{-\alpha}, \>
and the cumulative distribution is
\< P_\infty (x) &=&1- \left(  \frac{x}{x_\MIN} \right)^{1-\alpha}. \label{plinfcdf} \>

An useful property of the power law distribution is its relation to the exponential distribution.
By a logarithmic transformation the power law distribution becomes proportional to an exponential, $  x^{-\alpha} = e^{-\alpha \log_e x}$.
Due to this proportionality it is possible that some techniques for estimation and testing, which were  developed for the exponential distribution, can be used for the power law distribution.

\section{Estimating the parameters}\label{estimators}
There exist in the literature a variety of methods to estimate the parameters, exponent and upper limit, of a power law distribution.
In this first part of the paper we describe them and compare their properties.

\subsection{Methods}

\subsubsection{\label{binning} Binning}
A particularly simple method is to fit a linear relation to the data grouped in bins of constant size in logarithmic space.
As shown by \citet{maizapellaniz+ubeda2005} this method can yield biased results, i.e. results which systematically deviate from the actual parameter, and do not allow one to estimate a possible upper limit.

A solution to avoid biased results was given by \citet{maizapellaniz+ubeda2005}, which modified the binning scheme to a constant number of data points per bin and fitted the expected number instead of performing a linear regression.
In this way the estimate for the exponent, $\hat{\alpha}$, can be obtained, together with an estimate of its uncertainty, which is consistent with the sampling variance of $\hat{\alpha}$.
In extension of their work we investigate the properties of the estimate for the upper limit, derived from the normalisation constant of the frequency distribution, $\hat{k}= n ( 1-\alpha)/(x_\mathrm{MAX}^{1-\alpha} - x_\mathrm{MIN}^{1-\alpha} ) $, which is given by
\< \hat{x}_\MAX &=& \left( n \frac{1-\hat{\alpha}}{\hat{k}} + X_{(1)}^{1-\hat{\alpha}} \right)^{\frac{1}{1-\hat{\alpha}}}, \>
where the smallest observation ($X_{(1)}$) is used as an estimate for $x_\MIN$.

Not only the choice of constant-size or variable-size bins has influence on the results of binning, but also the number of bins.
\citet{dagostino+stephens1986} give as the optimal number of bins $2n^{2/5}$ for $n$ data points.
A smaller number of bins reduces the bias but increases the standard deviation \citep[cf. Table 1 to 3 of ][]{maizapellaniz+ubeda2005}.

\subsubsection{\label{ccdest} Complementary cumulative distribution function plot}
\begin{figure}
\includegraphics[width=8.5cm]{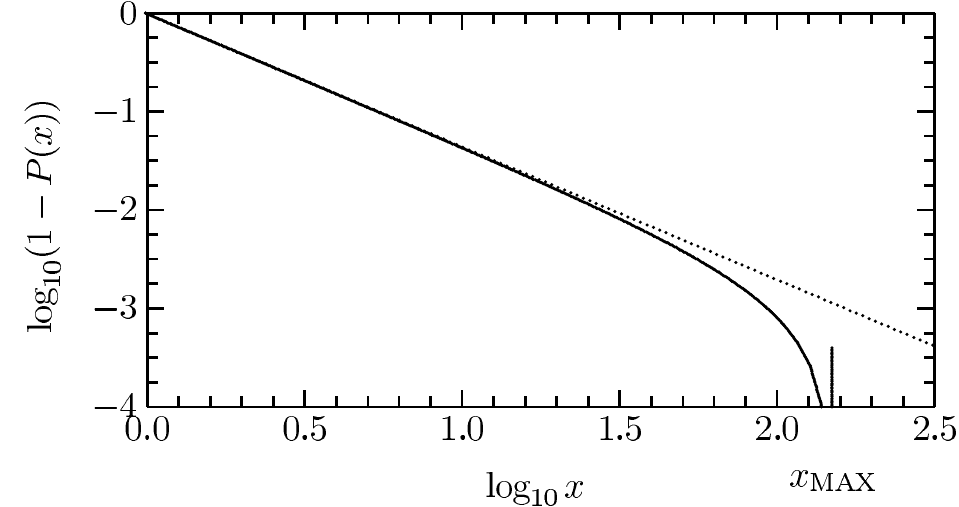}
\caption{\label{ccdfschema} Complementary cumulative DF (CCDF) plot for an infinite (dotted line) and truncated (solid line) power-law pdf ($\alpha=2.35$, $x_\MIN=1$, $x_\MAX=150$, shown by the vertical thick bar).
For the truncated case a characteristic turn-down appears at the upper end.}
\end{figure}

\citet{koen2006} presented a method to estimate both the exponent and the limits of a power law.
This method is based on a particular graphical representation of the data, the complementary cumulative DF (CCDF) plot (Fig. \ref{ccdfschema}). 
Data stemming from an infinite power law follow a linear relation with slope $1-\alpha$ in a plot of $\log (1 - P_\infty (x))$ versus $\log x$, as can be seen easily by taking the $\log$ of eq. \ref{plinfcdf}.
For a truncated power law, a turn-down appears at the high end.
Estimates for the exponent and limits are obtained by fitting $\log (1 - P(X_{(i)} ) )$ (with $P(x)$ from eq. \ref{pltrunccdf} and the ordered data $X_{(i)}$) to $\log (1 - \frac{i-0.5}{n})$, with using $\frac{i-0.5}{n}$ for the empirical cumulative distribution function.

\subsubsection{Beg's estimator}
A power-law distribution is closely related to the exponential distribution. 
Therefore it is possible to apply the uniformly minimum variance unbiased estimators for the slope and limits of a truncated exponential distribution, developed by \citet{beg1982,beg1983} to log-transformed power law data, as shown by \citet{beg1983}.
Although these estimators are theoretically an optimal solution, they are only partially practicable, since their computation is numerically difficult and impossible for large data sets.
We therefore developed a recursive form which is applicable to arbitrarily large data sets.
The original and recursive formulae are given in the Appendix.

\subsubsection{Maximum Likelihood (ML) estimator}
The Maximum Likelihood estimator for the exponent was given by \citet{crawford-etal1970}, who also included an upper limit.
In our case the upper limit is intrinsic to the distribution function, i.e. the upper limit needs to be estimated simultaneously with the exponent.
The likelihood function for a random sample of size $n$ from a truncated  power law DF is
\< \mathcal{L} &=& \prod_{i=1}^n p(x_i; \alpha; x_\MIN,x_\MAX)\\
&=& \left( \frac{1 -  \alpha}{x_\MAX^{1-\alpha}  -   x_\MIN^{1-\alpha}}\right)^n \prod_{i=1}^n x_i^{-\alpha}    \>
The estimator for the exponent is obtained by maximising the log-likelihood,
\< \log \mathcal{L} \!\!\!\!
 &=& \!\!\!\! n \log (1 \! - \! \alpha) \! - \! n \log ( x_\MAX^{1-\alpha} \! - \!  x_\MIN^{1-\alpha} ) - \alpha \sum_{i=1}^n \log x_i . \label{loglik}\>
The maximisation can be performed by finding the root of the derivative with respect to $\alpha$ of eq. \ref{loglik}.
The estimator $\hat{\alpha}_{ML} $ is then the solution of
\< -\frac{n}{1 -  \hat{\alpha}_{ML} } + n \frac{Z^{1  - \hat{\alpha}_{ML} } \log Z  -  Y^{1  -  \hat{\alpha}_{ML} } \log Y}{Z^{1-\hat{\alpha}_{ML} } - Y^{1-\hat{\alpha}_{ML} } } - T \!\!\!\!\! &=& \!\!\!\!\! 0, \>
with $Y= \min X_i$, $Z = \max X_i$ and $T=\sum_{i=1}^n \log X_i$.

The ML estimates for the upper  limit $\hat{X}_\MAX = \max X_i$ \citep[see e.g.][]{aban-etal2006}.
It is obvious that this estimate will be biased since the upper limit is larger than the largest data point.

\subsubsection{\label{modml} Bias-free estimators based on the maximum likelihood estimator}
It is possible to construct a minimum variance unbiased estimate of the exponent from the maximum likelihood estimate, as shown by \citet{crawford-etal1970} or \citet{baxter1980}.
For the infinite case  the ML estimator for the exponent is given by
\< \hat{\alpha} - 1 &=& \frac{n}{T-n \log_e Y}, \>
with $Y= \min X_i$ or the given lower limit, and $T=\sum_{i=1}^n \log_e X_i$.
The unbiased estimator  is then
\< \hat{\alpha}' -1  &=& \frac{n-1}{n} ( \hat{\alpha} -1) \label{baxterest}\>
(if both the exponent and the lower limit should be estimated then $(n-2)/n$ has to be used  \citep{baxter1980}).

\begin{figure}
\includegraphics[width=8.7cm]{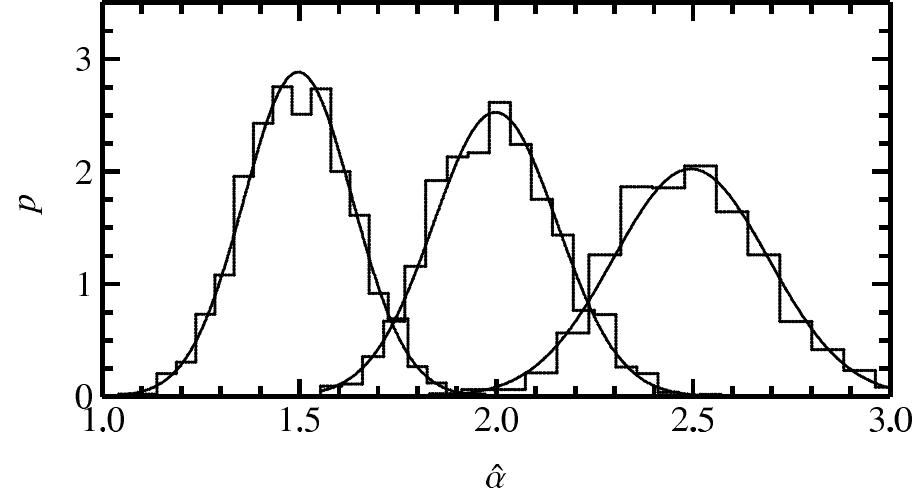}
\caption{\label{exponentmodml}
Distributions of the modified ML estimates for the exponent (histogram), derived from Monte-Carlo samples of size 1000 for three input values ($1.5$, $2.0$, and $2.5$).
They follow a Gaussian with mean and variance derived from the samples.
For larger exponents the variance increases.
}
\end{figure}

The simple relation between the ML estimator and the unbiased estimator for an infinite 
power law suggests a similar relation for the truncated case.
However, for a truncated power law a closed form of the ML estimator is not available.
This makes a proof of an unbiased estimator very difficult and maybe even impossible, since the distribution of the estimate cannot be calculated analytically.
Nevertheless, it is not unreasonable to assume that a simple modification of the ML estimate also leads to unbiased results.
A different pre-factor, depending only on the number of data, should give the expected result.
We found that
\< \hat{\alpha}_\MML -1 &=& \frac{n}{n-2} (\hat{\alpha}_{ML} -1 ). \label{modmlalpha}\>
(MML = Modified Maximum Likelihood) provides quasi-bias-free estimates.
The pre-factors $\frac{n-2}{n}$ of \citet{baxter1980} or $\frac{n-3}{n}$ ($n-3$ because there is an additional parameter, the upper limit) lead to biased results.
The distribution of the exponents estimated using this method follows a Gaussian, as can be seen in Fig. \ref{exponentmodml}, with an increasing variance for an increasing exponent.

The bias of the ML estimate for the upper limit can also be significantly reduced by appropriate modifications.
\citet{hannon+dahiya1999} developed such a modified estimator for the exponential distribution.
This estimator can also be used for the power law distribution and takes then the form (with the ML estimate of the exponent replaced with the bias-reduced form)
\< \hat{x}_\MAX &=& X_{(n)} \left( 1 + \frac{ e^G -1 }{n} \right)^{ \frac{1}{ 1-\hat{\alpha}_\MML } }
\label{modmlmmax}, \>
with
\< G &=& ( 1-\hat{\alpha}_\MML ) 
\log_e \left( \frac{ X_{(n)} }{ X_{(1)} } \right), \> 
where $X_{(1)}$ is the smallest and $X_{(n)}$ is the largest data point.
The properties of the modified estimate are discussed in the next Section.

\subsection{Performance of the estimators}\label{estimatorresults}
\begin{figure*}
\includegraphics[height=8cm]{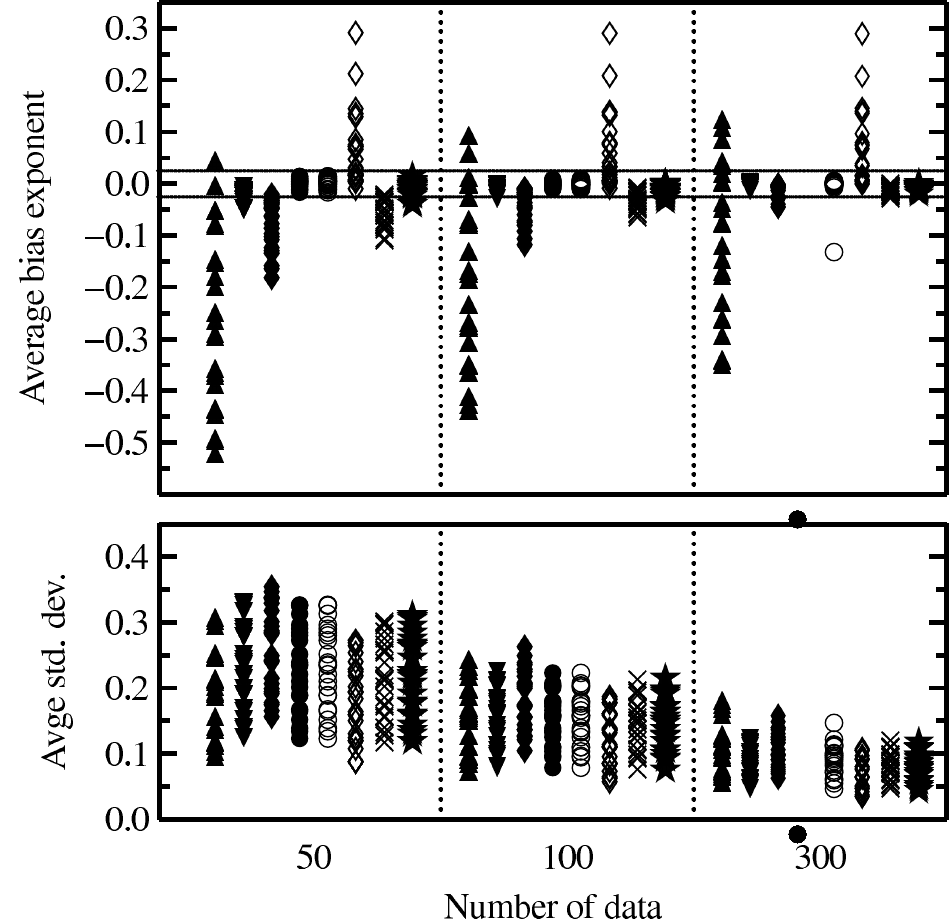}
\hspace{2em}
\includegraphics[height=8cm]{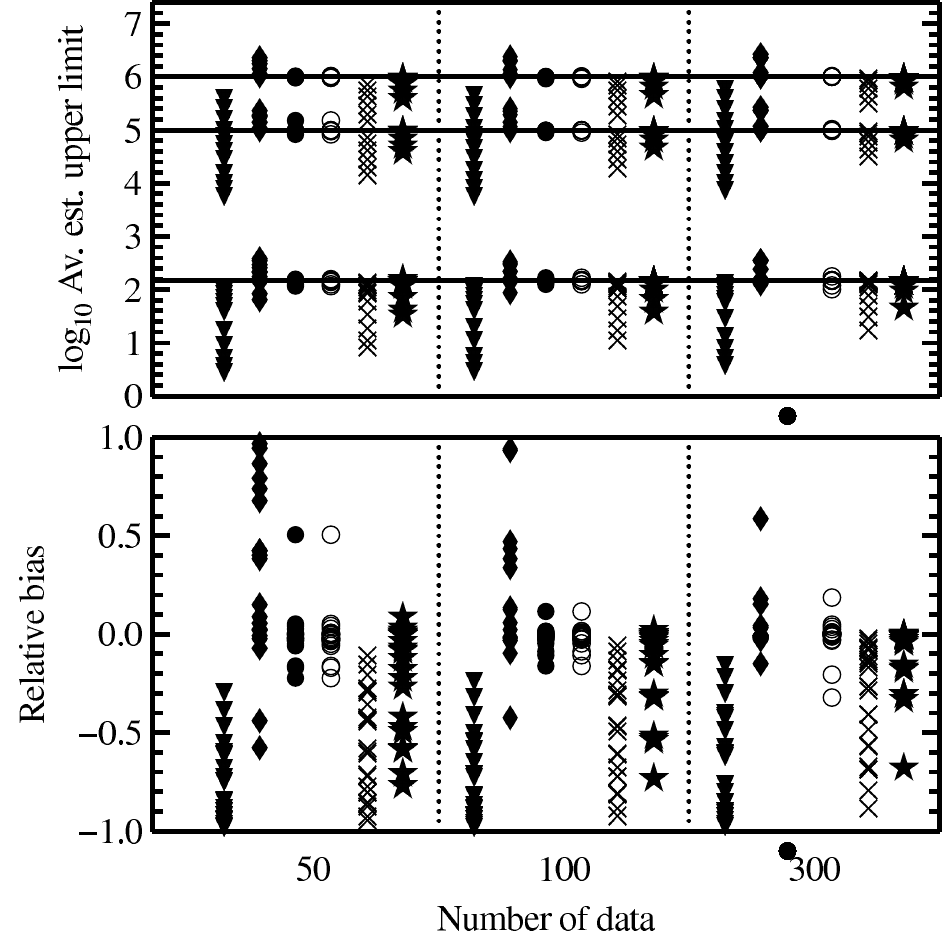}
\caption{\label{betacomp}\label{variancecomp}\label{mmaxcomp}\label{mmaxcomprel}
In the left panel the results for the exponent are shown, on top the average bias (calculated using eq. \ref{bias}, the horizontal lines mark $\pm0.025$) and below the average standard deviation (eq. \ref{stddev}) of the estimated exponents.
In the right panel, average estimates of the upper limit (lines mark the true values) are displayed in the upper part and below the average relative bias is shown (eq. \ref{relbias}).
The parameter combinations are given in the text, Sec. \ref{estimatorresults}.
The symbols refer to:
$\blacktriangle$ constant size binning using linear regression;
$\blacktriangledown$ variable size binning using $\chi^2$;
$\blacklozenge$ CCDF plot fitting.
$\lozenge$ ML estimator without including truncation;
$\bullet$ Beg's estimator;
(This estimator starts to fail for $n\ge150$, dots below the x-axis indicate a failed experiment);
$\circ$ Beg's estimator in the recursive form;
$\times$ ML estimator;
$\bigstar$ Modified ML estimator;
}
\end{figure*}

After introducing a number of methods of estimation, we compare their properties.
The quality and usability of an estimator is determined by several factors.
A main demand is that an estimate is on average equal to the actual parameter, i.e. bias-free.
Also, the variance of the estimate should be as small as possible and it should be numerically robust.

To study these properties we carried out  a  set of Monte-Carlo experiments, each of size 1000, with parameters in the typical range of astronomical applications.
The values for the exponent range from $1.6$ to $2.85$ in steps of $0.25$.
For each exponent four pairs of limits were used ($\{0.5,150\}$ and $\{10,150\}$ corresponding to the stellar mass function, $\{10^3,10^5\}$ and $\{10^4,10^6\}$ corresponding to the mass function of young star clusters).
The last varied parameter was the number of data ($50$, $100$, $300$).
For the binning methods the number of bins was chosen according to \citet{dagostino+stephens1986} ($2n^{2/5}$), which gave $9$, $12$ and $19$ bins, respectively.

As diagnostics for the performance in estimating the exponent we choose the average bias of an estimator for a given parameter set,
\< \bar{B}(\alpha ) &=&  \frac{1}{1000} \sum_{i=1}^{1000}  \big( \hat{\alpha}_{i} - \alpha  \big), \label{bias} \>
and the standard deviation,  
\< \bar{S} (\alpha) &=& \sqrt{ \frac{1}{1000} \sum_{i=1}^{1000} \big( \hat{\alpha}_i - \bar{\alpha} \big)^2 } \label{stddev}.\>

The left panel of Figure \ref{betacomp} summarises the results for the bias.
Two horizontal lines at $\pm 0.025$ embrace the region, in which we consider the bias as negligible.
The general trend is that the bias decreases with an increasing number of data (except for the ML estimator wich does not include a truncation).
The corresponding results for the standard deviation also decrease with larger size of the data set.

Variable-size binning gives effectively bias-free exponents for samples having a moderate size or larger.
The method of Koen is biased towards lower exponents.
The results from Beg's estimator are very good, but the method fails for large data sets, even in the recursive form. 
A maximum likelihood estimate without considering the truncation can lead to a significantly overestimated exponent.
But when the truncation is considered, the bias is small and effectively vanishes if our modified version is used.

\begin{figure}
\includegraphics[width=8.7cm]{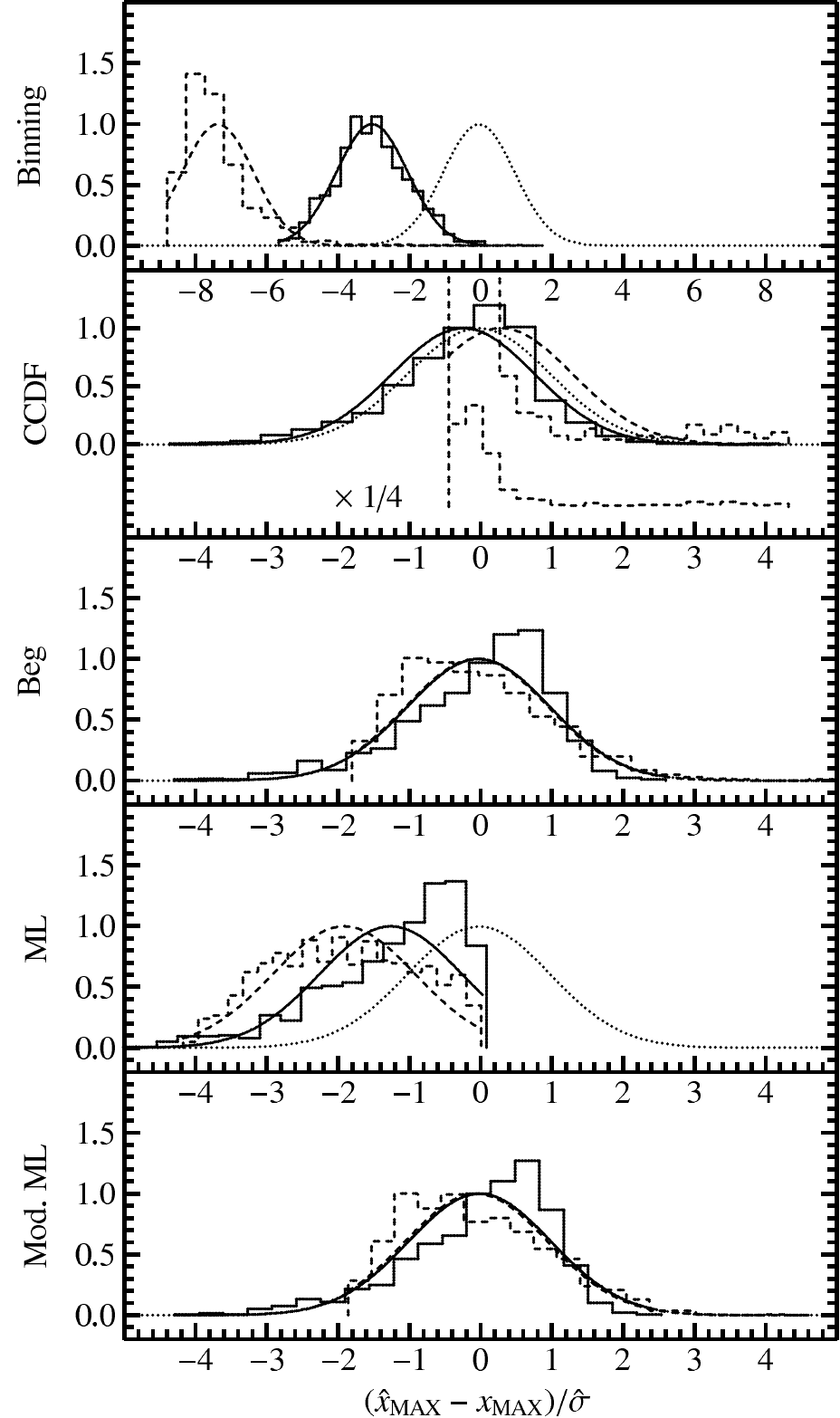}
\caption{\label{upperlimithist}
Distribution of the estimated upper limits for the different methods (histograms; parameters: $\alpha=2.0$; $n=100$; solid: limits $\{10,150\}$, dashed: limits $\{1000,100000\}$.)
The x-axis has been scaled such that a unbiased estimate should follow a Gaussian of zero mean and unit variance (dotted).
Also shown is a Gaussian with mean and variance derived from the estimates.
}
\end{figure}

The results for the  estimates of the upper limit are
shown in the right part of  Fig. \ref{mmaxcomp}.
Generally it can be observed that a larger upper limit leads to larger absolute deviations in the estimate.
Because the upper limits used in this study span a wide range of values it is not convenient to compare the absolute biases as for the exponents.
The relative bias (also displayed in Fig. \ref{mmaxcomprel}),
\< \tilde{B} (x_\MAX) &=& \frac{1}{1000} \sum_{i=1}^{1000} \left( \frac{\hat{x}_{\MAX,i} - x_\MAX}{x_\MAX} \right) , \label{relbias} \>
is a better measure of trends.
Furthermore the normalised distributions of the estimates are shown in Fig. \ref{upperlimithist},  for two parameter sets ($\alpha=2.0$,  $n=100$, limits $\{10,150\}$ and $\{1000,100000\}$).
The histogram and a Gaussian with mean and variance ($\sigma$) calculated from the Monte-Carlo sample are rescaled by
$x' = \frac{x-x_\MAX}{\sigma} $
and the y-axis is scaled such that the peak of the Gaussian is 1.
If  the estimator is not biased and can be approximated by a Gaussian, then the normalised distribution should follow a Gaussian with zero mean and unit variance.

The upper limit is underestimated by using the normalisation constant of the variable-size binning method.
The results of fitting the CCDF plot are peaked around the input value, but can have a long tail of very high estimates (for the limits  $\{1000,100000\}$).
If in the CCDF plot the data show no strong curvature at the upper end, then the estimated upper limit is very large.
The distribution of estimates obtained with Beg's estimator are in reasonable agreement with a Gaussian, but not completely symmetric around the mean.
If the largest data point is used (i.e. the direct ML estimate), then the upper limit is underestimated, with a distribution limited by the actual value.
With the modification (eq. \ref{modmlmmax}) the distribution becomes similar to the one of Beg's estimator, spreading around the true value.
Although not completely symmetric around the mean it can be sufficiently approximated by a Gaussian, and has no outliers as fitting the CCDF plot.

In summary, when both the exponent and the upper limit of a truncated power law should be estimated, our modified ML method performs best in terms of bias and stability, being similar to Beg's uniformly minimum variance unbiased estimator, but without the numerical instability.
gives the best results in terms of bias and stability.

\section{Is a power law consistent with the data?}\label{secgoodness}
For a thorough analysis of data which are assumed to stem from a power-law distribution it is not sufficient just to estimate the parameters.
The parameter estimation answers the question which power-law fits the data best, but leaves open whether the data are originating from a power law \textit{at all}.
Or to put it differently: is the (truncated) power law a good  parent distribution function of the data?
The need to answer this question has already been stressed by \citet{crawford-etal1970}.
This question can be addressed by a graphical inspection of the data, which is discussed in the next Section.
After the informal visual methods more objective goodness-of-fit techniques are 
discussed.

\subsection{Graphical inspection of the data}\label{plottingmethods}
\begin{figure*}
\includegraphics[width=8.5cm]{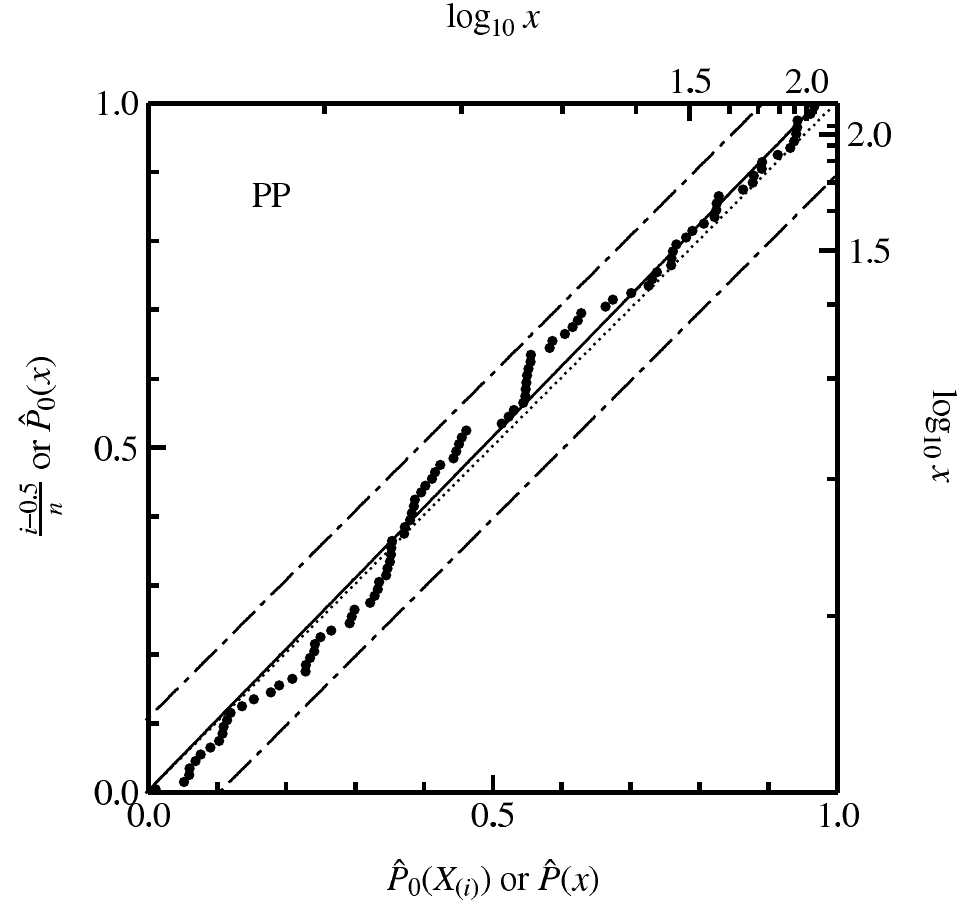}
\hspace{2em}
\includegraphics[width=8.5cm]{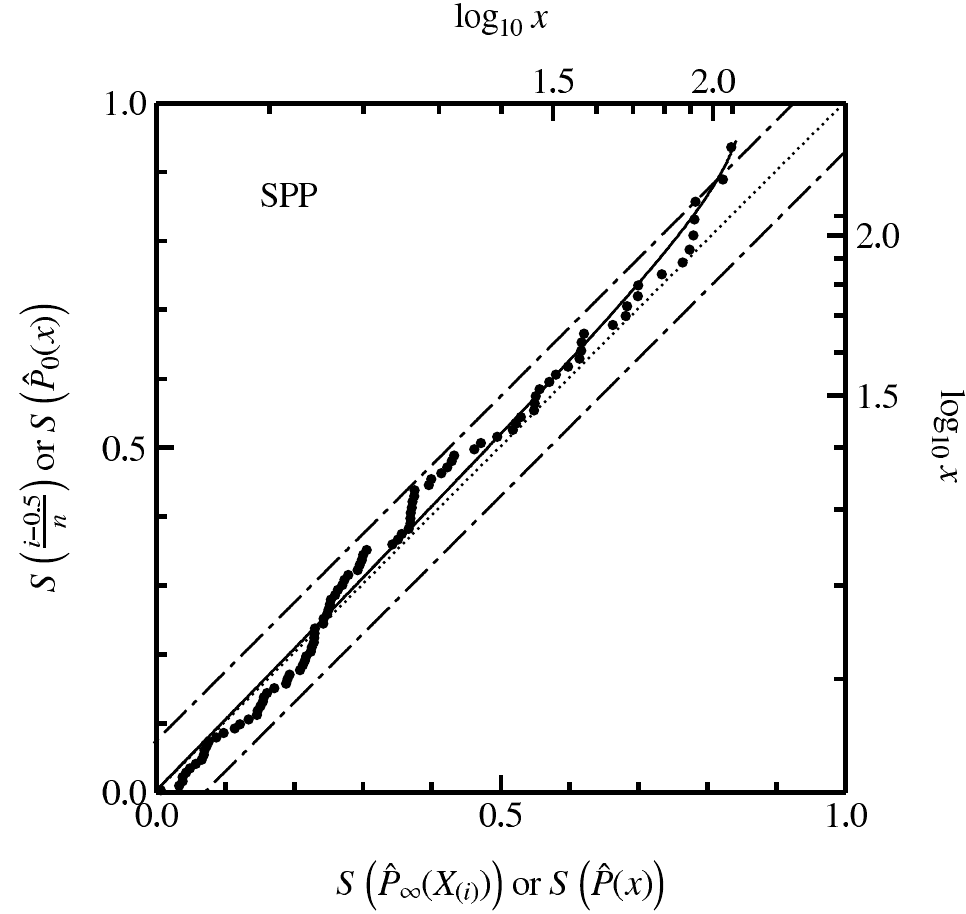}
\caption{\label{ppplot}\label{sppplot}
Example for a percentile-percentile plot (PP, left) and a stabilised percentile-percentile plot (SPP, right) for the null hypothesis of an infinite power law (=diagonal), using 100 data points sampled from a truncated power law ($\alpha=2.35$, $x_\MIN=1$ and $x_\MAX=150$).
Also shown are the curves for a truncated power law (solid line, parameters as estimated, $\hat{\alpha}=2.35$ and $\hat{x}_\MAX=149$). 
The acceptance region of the  (in the right plot stabilised) Kolmogorov-Smirnov statistic (significance level 5\%) is given by the two  parallels to the diagonal.
The data lie within this region in the PP plot, wherefore from the PP plot the infinite power law cannot be significantly rejected.
After stabilisation the KS test is more powerful and thus allows us to detect truncation in contrast to the PP plot.
}
\end{figure*}

A common approach to find the parent DF of a data set is to use a histogram as a non-parametric estimate of the form of the parent DF.
If in a logarithmic plot the histogram of e.g. stellar masses is a straight line, a power-law is usually assumed as the parent DF.
However, a power-law is a heavy-tailed distribution and has only a few counts per bin in the tail.
Thus the scatter in a histogram is large in the upper regime and makes deviations from a power-law hard to detect.
Alternative, heavy-tailed distributions lead to nearly indistinguishable histograms.
It is for example not possible to decide whether the power-law is truncated or not.
Therefore a histogram only allows us to roughly determine the parent DF.

A display of the data which avoids grouping them into cells is the probability-probability (or percentile-percenile, PP) plot.
For the PP plot the data have first to be sorted in ascending order, $X_{(i)} < X_{(i+1)}$.
The x-values then follow as the ``theoretical'' percentiles, which are the values of the cumulative DF for the i-th data point, $P(X_{(i)},\hat{\alpha},\hat{x}_\MIN,\hat{x}_\MAX)$, calculated with the estimated parameters.
As y-values the ``empirical'' percentiles are used, given by $\frac{i-0.5}{n}$.
Both axes range from $0$ to $1$, and when the data lie on the diagonal then they agree with the null-hypothesis.
Hypotheses other than the null hypothesis can be shown by plotting the pairs $\{$x=alternative cumulative DF,y=null cumulative DF$\}$.
An example for a PP plot with an infinite power law as null hypothesis (diagonal) is shown in Fig. \ref{ppplot}, but with data generated from a truncated power law.
The curve for the alternative hypothesis of a truncated power law (of the same exponent, solid line) barely deviates from the diagonal and does not allows one to distinguish the infinite and truncated versions.
In this plot the acceptance region of the Kolmogorov-Smirnov (KS) test can be directly shown as parallels to the diagonal, in Fig. \ref{ppplot} calculated with a significance level of 5\%.
The data do not exceed this region, not even in the tails, giving evidence for the known insensitivity in the tails of the Kolmogorov-Smirnov statistic.

Before showing a way to improve the insensitivity of the PP plot in the tail we shortly compare it with other possible plots which show all available data (see e.g.  \citet{chambers-etal1983} or \citet{wilk+gnanadesikan1968}).
A plot which uses e.g. the $X_{(i)}$ as x-values (the ``empirical cumulative density plot'') has a curved reference line for both an infinite and truncated power law, with only small and not well perceptible differences between them.
Following the suggestion of \citet{koen2006}, in a plot of $\log (1-P(X))$ against $\log X$ (the CCDF plot, see Fig. \ref{ccdfschema} and Sec. \ref{ccdest}) the data should only show a curvature for a truncated power law.
But from the results for the estimator based on this plot, the scatter in the highest data points can be large, and a graphical goodness-of-fit criterion would not be very sensitive.
A third alternative to the PP plot would be a plot of the inverse cumulative DF ($P^{-1} (\frac{i-0.5}{n})$, giving the expected value for the data point $X_{(i)}$) against the ordered data $X_{(i)}$ (``quantile-quantile'' plot).
This plot has a linear reference, the upper tail would be curved when constructed for the infinite null hypothesis but with truncated data.
Again, as for the complementary cumulative density plot there is large scatter, and
additionally the inverse cumulative DF has to be calculated.

\subsection{The stabilising transformation and the SPP plot}\label{secstab}
Goodness-of-fit methods based on the empirical cumulative DF, such as the PP plot or the KS test, have the advantage that their intrinsic properties do not depend on the actual choice of the null hypothesis.
A PP-plot for e.g. Gaussian variates looks identical (modulo the random scatter) to a PP plot for power-law data.
The reason for this is that by taking the cumulative DF the data are transformed to uniformly distributed variates, if they are following the null hypothesis.
Therefore the location and variances of the points in the PP plot are independent of the null hypothesis distribution, as is the distribution of the KS statistic.
This transformation reduces the goodness-of-fit task from arbitrary DFs to testing for uniformity.
However, the variances of uniform ordered variates are not independent from their position: the scatter in the PP plot is larger in the middle than in the tails.
Thus a test which measures the differences between the expectation and the empirical values of uniformly distributed ordered data will be dominated by the points with the larger variances.
Hence the insensitivity of the KS-test in the tails.

A way to overcome the unequal variances was introduced by \citet{michael1983}.
The stabilising transformation of uniform variates $u$ (= the cumulative DF),
\< S_0 (u) &=& \frac{2}{\pi} \arcsin (\sqrt{u})  \label{stabtrans0} \>
gives asymptotically equal variances of the transformed ordered variates.
In a stabilised PP (SPP) plot every part of the plot has the same weight and no region is particularly emphasised.
Although the distribution of the $S_0(u)$ is not uniform any more, tests based on the differences between expectation and empirical value can still be used (as long as they do not use other properties of the uniform distribution).
These transformed tests are equally sensitive to every part of the distribution function.

However, for testing the tail-behaviour of a DF it is useful to emphasise the tail.
This can be achieved by using only a half transformation, which is possible because $S_0$ is symmetric around the point $\{0.5,0.5\}$ and the interval $[0.5,1]$ is mapped onto $[0.5,1]$.
A one-sided transformation of the percentiles to stabilise a right-tailed distribution consists then of three steps.
First, map the interval $[0,1]$ on $[0.5,1]$, then use $S_0$, and lastly map $[0.5,1]$ back on $[0,1]$.
The formula for this is
\< S (u) &=& 2 S_0 ( 0.5 + 0.5u) -1.\label{stabtrans} \>
We use $S$ instead of $S_0$ in the SPP plot (Fig. \ref{sppplot}) and in the goodness-of-fit tests which are related to it, because of the one-tailed power law distribution.

\subsection{Goodness-of-fit Tests}\label{tests}
A goodness-of-fit test provides an objective way to ``measure'' the agreement of the fit with the data.
We follow here the Neyman-Pearson ansatz of hypothesis tests.
At first the type I error probability or significance level needs to be specified.
This is the rate at which the test is allowed to falsely declare a data set as too discrepant to be compatible with the null hypothesis (the assumed distribution function), even though it is in reality consistent.
There is a value of the distribution of the test statistic, the critical value, which corresponds to this rate.
If the value of the test statistic calculated from the data set then exceeds the critical value the null hypothesis is rejected for the data set.

For some tests, the distribution of the test statistic can be calculated analytically for a fully specified null hypothesis, i.e. if no parameter is estimated.  
If parameters are estimated, the distribution of the test statistic is not universal any more, but depends on the properties of the specific estimator.
The distribution of the test statistic can then be obtained using a Monte-Carlo approach, of which follow the critical values.
Typically the such derived critical values are larger than for a fully specified hypothesis \citep[cf.][for the Kolmogorov-Smirnov test for the normal and exponential distribution]{lilliefors1967,lilliefors1969}.
Therefore, if the critical values for the fully specified hypothesis are used when parameters are estimated, the results are conservative with an actually smaller type I error, but also less powerful.

The significance level is not the only quantity characterising a statistical test.
It can happen that a data set is not too discrepant to be rejected, but actually does not stem from the null hypothesis, i.e. a type II error occurs.
The probability that a type II error does not occur is the (statistical) power of the test.
If the power of the test is small then it is not very selective and the alternative hypothesis cannot be strongly excluded.
For a given test the power can differ for various alternative hypotheses of the distribution.
A demand for a general purpose test is to be powerful against a wide variety of alternative hypotheses.

In order to be able to evaluate the ``strength'' of a statement concluded from a statistical test, it is therefore necessary to know the type I and type II error rates (the significance level and the power).
However, not every test has the same power, therefore we  conduct in what follows a power study which has the purpose of finding a powerful goodness-of-fit test to decide between infinite and truncated power laws.
The astrophysical motivation for the choice of these hypotheses is the discussion in the literature about an upper mass limit for the distribution of stars in a star cluster \citep[cf.][]{weidner+kroupa2004,oey+clarke2005,koen2006} or about an upper limit for the star cluster luminosity function \citep[cf.][]{gieles-etal2006b}.

\subsection{Description of the goodness-of-fit test statistics}
Goodness-of-fit tests can roughly be classified as tests based on the empirical DF (EDF), based on distance measures (e.g. the KS test) or the correlation coefficient, tests especially developed for a chosen null hypothesis (e.g. the Shapiro-Wilk test for exponentiality), and tests to distinguish between two hypotheses (e.g. the Likelihood Ratio).
Below we describe the test statistics used for a comparison. 

For the selection of the tests included in the comparison the properties of tests for exponentiality can be used, which can be found in the studies of \citet{stephens1978}, \citet{dagostino+stephens1986} and \citet{gan+koehler1990}.
\citet{gan+koehler1990} which use EDF based tests included the alternative hypothesis of a truncated exponential, finding only very low powers.
With the stabilising transformation \citet{kimber1985} finds a larger power for the KS test, but did not include the truncated alternative.
We include some EDF based tests in the original and stabilised version, as well as some tests based on tests for exponentiality, and two tests explicitly for truncation.

\begin{table*}
\caption{\label{powertab} Results of the power study. Tests are conducted under the null hypothesis of an infinite power law against the alternative hypothesis of a truncated power law with a type I error level of $\alpha_I=0.05$. The first column gives the size of the data set, $n$, and the second column the value of the exponent, $\alpha$. The other columns give the power of the test statistic indicated in the top row in percent. 
The power is the fraction of the time in which the data drawn from the alternative hypothesis would be rejected (at the 5\% level) as coming from the null hypothesis.
The numbers on top of each group are the lower and upper limit of the parent distribution function.}
\input{power.tex}
\end{table*}

\subsubsection{EDF statistics based on distance measures}
The most prominent goodness-of-fit test is the \textit{Kolmogorov-Smirnov (KS) statistic} \citep{stephens1978,dagostino+stephens1986,gan+koehler1990},
\< D &=& \max_{1 \le i \le n} \left| \frac{i-0.5}{n} - \hat{P}_{(i)} \right| + \frac{1}{2n}, \>
which is the largest vertical distance between the data and the diagonal in the PP plot (Rejection for $D > D_\crit$).
The largest distance can also be measured in the stabilised PP plot, leading to the 
\textit{stabilized Kolmogorov-Smirnov statistic} \citep{michael1983,kimber1985},
\< SD &=& \max_{1 \le i \le n}   \left| S\left(\frac{i-0.5}{n}\right) - S(\hat{P}_{(i)}) \right|, \>
\citet{kimber1985} found that for the exponential distribution these statistics are more powerful than the originals, but used a complete stabilising transformation (for both tails, $S_0$, eq. \ref{stabtrans0}). 
Here only a right-tail-stabilising transformation ($S$, eq. \ref{stabtrans}) is used, since it is more appropriate for the right-tailed power law and gives a better power.

Another ``measure of discrepancy'' is the sum of the squared distances from the diagonal to the data point in a PP plot, the \textit{Cram\'er-von Mises statistic} \citep{anderson+darling1952,stephens1978,dagostino+stephens1986,gan+koehler1990},
\< C^2 &=& \sum_{i=1}^n \left( \hat{P}_{(i)} - \frac{2i-1}{2n} \right)^2 - \frac{1}{12 n}.\>
Like the KS statistic, this measure can be  used in the stabilised PP plot,
yielding the new \textit{stabilised Cram\'er-von Mises statistic},
\< SC^2 &=& \sum_{i=1}^n \left( S\hat{P}_{(i)} - S\left( \frac{2i-1}{2n} \right) \right)^2. \>
A modified form of the Cram\'er-von Mises statistic is the \textit{Anderson-Darling statistic},
\citep{anderson+darling1952,stephens1978,dagostino+stephens1986,gan+koehler1990}
\< A^2 &=&\!\!\! - \sum_{i=1}^n \frac{(2i\!\!-\!\! 1)}{n} \!\! \big( \log_e (\hat{P}_{(i)})\!\! -\!\! \log_e (1-\hat{P}_{(n+1-i)})  \big)\!\! -\!\! n, \>
which  gives more weight to the tails of the distribution.

\subsubsection{EDF statistics based on the correlation coefficient}
The correlation coefficient is a measure for linearity, given by
\< R^2 (X,Y) &=& 
\frac{
\big( \sum_{i=1}^n (X_i - \bar{X}) (Y_i - \bar{Y}) \big)^2 
}{
\sum_{i=1}^n (X_i - \bar{X})^2 \sum_{i-1}^n (Y_i -\bar{Y})^2 
}. \label{corrcoeff} \>
For perfect linearity $R^2$ has the value 1 or $-1$ and for uncorrelated points $R^2=0$.
If the points $\{X,Y\}$ are always positive as in our cases then $R^2$ lies in the interval $[0,1]$. The rejection criterion is $R^2 < R^2_\crit$.

The correlation coefficient can for example be used in the quantile-quantile plot,
\< r^2 &=& R^2 \left( X_{(i)} ,\hat{P}^{-1}_{(i)} \right) , \>
but has, as the quantile-quantile plot, the disadvantage of needing the inverse distribution function.

Another possibility is to use the correlation coefficient in the PP plot (\textit{PP correlation statistic}; \citealp{gan+koehler1990}),
\< k^2 &=& R^2 \left( \hat{P}_{(i)}, \frac{i-0.5}{n} \right),  \>
or in the stabilised PP plot (\textit{stabilised PP correlation statistic}, first proposed here),
\< Sk^2 &=& R^2 \left( S\hat{P}_{(i)}, S \left( \frac{i-0.5}{n} \right) \right) .\>
The PP correlation statistic can be modified, as suggested by \citet{gan+koehler1990}, to force the points to go through $\{0.5,0.5\}$.
This is done by replacing in eq. \ref{corrcoeff} $\bar{X}$ and $\bar{Y}$  by 0.5, denoting the modified version $R_0^2$.
\citet{gan+koehler1990} found that the {\it modified PP correlation statistic},
\< k_0^2 &=& R_0^2 \left( \hat{P}_{(i)}, \frac{i-0.5}{n} \right) , \>
 is somewhat more powerful than $k^2$.
Again, the analogous procedere is possible in the stabilised PP plot, giving the {\it stabilised modified PP correlation statistic},
\< Sk_0^2 &=& R_0^2 \left( S\hat{P}_{(i)}, S \left( \frac{i-0.5}{n} \right) \right) . \>

\subsubsection{Statistics based on tests for exponentiality}
Due to the connection of the power law and exponential distribution, the various tests for exponentiality available in the literature are applicable for our purposes \citep[cf. e.g.][]{beirlant-etal}.
Since there is only a proportionality between $p(x)$ and $p_e (\log_e x)$ most of the derived (exponential) null distributions for the following statistics are no longer valid for an infinite power law.

The \textit{Shapiro-Wilk statistic} \citep{shapiro+wilk1972,stephens1978,dagostino+stephens1986},
\< W &=& \frac{n}{n-1}(\bar{X'} - X_{(1)}' ) \sum_{i=1}^n (X_i' - \bar{X'})^2, \>
($ X_i' = \log_e X_i$)
is originally a two-sided statistic with minimum $(n-1)^{-2}$ and maximum 1 for the exponential case.
For the use with a power law the rejection criterion for the alternative hypothesis of a truncated power law distribution is $W>W_\crit$ in one-sided use.

The \textit{Jackson statistic} \citep{jackson1967,stephens1978,dagostino+stephens1986,beirlant-etal},
\< T &=& \frac{\sum_{i=1}^n t_{i,n} X_{(i)}'' }{\sum_{i=1}^n X_i''}, \>
with 
$ X_i'' = \log_e  (X_i/\hat{x}_\MIN  )$ and 
 $ t_{i,n} = \sum_{j=1}^i \frac{1}{n-j+1} $, is primarily the product of the ordered data and their expectation values $\lambda E(X_{(i)})=t_{i,n}$, comparable to correlation type statistics. 
The division by $\sum_{i=1}^n X_i$ removes the dependence on the scale parameter $\lambda$.
For a truncated power law alternative in one-sided use the rejection criterion is $-T > -T_\crit$.

Other tests for exponentiality are the the statistics of \citet{brain+shapiro1983}, the Moran statistic \citep{stephens1978,dagostino+stephens1986}, and the Greenwood statistic \citep{bartholomew1957,stephens1978,dagostino+stephens1986}.
We have also tested their powers when used for a power law, but they are not more powerful than the Shapiro-Wilk or Jackson statistic, and so we do not include details on them here.

\subsection{Tests for truncation}
The above described tests only allow one to distinguish whether the data are described by the null hypothesis or not.
When the null hypothesis is rejected the test has to be made again, now with the alternative hypothesis as new null hypothesis.
The \textit{Likelihood Ratio} test combines this two-stage procedure into one test, by which can be decided whether of the two hypotheses is favourable. 
We use here the Likelihood ratio in the same way as the test statistics above.
The test statistic is given by
\< \Lambda &=& \frac{\prod_{i=1}^n p_\infty(X_i;\hat{\alpha}_\infty)
}{\prod_{i=1}^n p(X_i;\hat{\alpha},\hat{x}_\MAX)} . \label{likelihoodratio}
\>
For the infinite case we used  the ML estimate which does not include a truncation to estimate the exponent and for the truncated case the modified ML for the exponent and upper limit.
For numerical reasons the logarithm of eq. \ref{likelihoodratio} is evaluated.

An answer to the problem of estimating parameter and simultaneously deciding between hypotheses can also be given in the Bayesian framework of statistics which is, however, not in the scope of this article.

A further specific test for  truncation is the \textit{exceedance statistic},
\< X &=& \max_{i \le 1 \le n} X_i \>
(the largest data point), is only designed to test whether the distribution function is truncated or not.
It cannot be used to detect a deviation from the power-law assumption.
Furthermore it is one-sided with rejection criterion $-X > -X_\crit$.

\subsection{Power comparisons}
The power of the various statistics was calculated at a significance/type I error level of $\alpha_I =0.05$ with parameters for the power law as given in Table \ref{powertab}.
The critical points for the null hypothesis of an infinite power law were calculated as follows.
For each of the parameter combinations, but with $x_\MAX=\infty$, a Monte-Carlo sample containing 1000 data sets was generated.
For each data set the parameters were estimated using the modified ML estimator (exponent eq. \ref{modmlalpha}, upper limit eq. \ref{modmlmmax}).
Then the test statistics were calculated using the estimates when necessary.
This gives the distribution of the respective test statistic, from which the critical value follows as the 95\% quantile.

For the power again a sample of 1000 data sets was generated, but now from a truncated power law.
As before the parameters were estimated and the statistics calculated.
The power is then the percentage of data sets with a test statistic smaller than the critical value.

The obtained powers are shown in Table \ref{powertab}.
The exceedance statistic, $X$, is the most powerful test for truncation.
However, it cannot be used for detecting deviations from the power law distribution.
Thus it has to be used in conjunction with one or more of the other tests which include a test for the power law family as the parent distribution function.

A general effect appearing for all statistics is that the power decreases with increasing slope and range of the limits.
By such changes the truncated distribution becomes -- informally speaking -- more similar to the infinite distribution and thus harder to discriminate.
Above $\alpha=2$ the performance of the tests drop significantly and therefore strong statements on truncation can barely be made.
Unfortunately for the Salpeter value of the slope, $\alpha=2.35$, the studied tests are mostly not powerful enough to decide whether an upper truncation is present or not.
However, in some not so extreme real cases such as the data set of massive stars in R136 a sufficient power can be achieved.
Furthermore, even if a truncation cannot be detected then deviations from a power law might still be discoverable.

Besides the general performance behaviour of the test statistics a further, rather surprising trend exists in the power.
The most powerful tests are not necessarily the tests derived especially for the power law distribution from tests for exponentiality. 
The stabilising transformation (eq. \ref{stabtrans}) strongly enhances the power of general-purpose ECDF or correlation statistics so that they outperform the specialised tests.
The Kolmogorov-Smirnov statistic which is known to be not very powerful \citep[cf.][]{gan+koehler1990} becomes, after stabilisation, more powerful than all other tests except for the exceedance test.
In their not stabilised forms the general-purpose tests are, as expected, less powerful than the specialised tests.
This enhanced power is a useful property since general-purpose statistics can easily be modified to tests for a different null hypothesis, e.g. a two-part power law.

In summary, the best test for truncation is the exceedance test, $X$.
To confirm the hypothesis of a power law and for better significance this test should be followed by some of the most powerful remaining tests.
These are, loosely ordered in descending power, the stabilised Kolmogorov-Smirnov test $SD$, the stabilised PP correlation test $Sk_0^2$, the stabilised Cram\'er-von Mises test $SC^2$ and the Jackson statistic $T$.

When a truncation is detected, then the hypothesis of a truncated power law has to be confirmed by again applying the respective statistics with this distribution (the truncated power law) as the null hypothesis.

\section{Examples}\label{examples}
\subsection{The massive stars in R136}
\begin{table}
\caption{\label{r136results} Estimates for the 29 most massive stars in R136. The standard deviations of the estimators were calculated using a Monte-Carlo sample of size 10 000. For the binning methods 5 bins are used. The bias was calculated using the results of the modified ML method as input values.
}
{\bf Data using \citet{chlebowski+garmany1991}}\\
\begin{tabular}{llllll}
\hline
Estimate for			& Slope $\hat{\alpha}$	& Bias slope	& $\hat{M}_\MAX$	& Bias $\hat{M}_\MAX$	\\
					&				&			& $[\Msun]$ 	&				\\
Const. Bins, LR			& 3.38$\pm$0.72	& 0.43		&			&				\\ Var. Bins, $\chi^2$		& 2.42$\pm$0.75	& -0.01		& 134$\pm$12	& -11				\\
CCDF				& 2.02$\pm$0.88 	& -0.11		& 140$\pm$9	& -1.9			\\ 
Beg					& 2.17$\pm$0.77	& $<$0.01		& 142$\pm$8 	& -0.4			\\
Beg, recursive 			& 2.17$\pm$0.77	& 0.01		& 142$\pm$8	& -0.4			\\
ML$\infty$				& 3.51$\pm$0.35      & 1.34               \\
ML					& 2.11$\pm$0.73 	& -0.06		& 136$\pm$7	& -8				\\
Mod. ML				& 2.20$\pm$0.78 	& 0.02		& 143$\pm$9	& $<$ 0.1			\\
\hline
\multicolumn{4}{l}{Results of \citet{koen2006} }\\
CCDF				& 2.10			&			& 143.9 \\
ML					& 2.11			&			& 136 \\ \hline
\end{tabular}
\end{table}

\begin{figure}
\includegraphics[width=8.5cm]{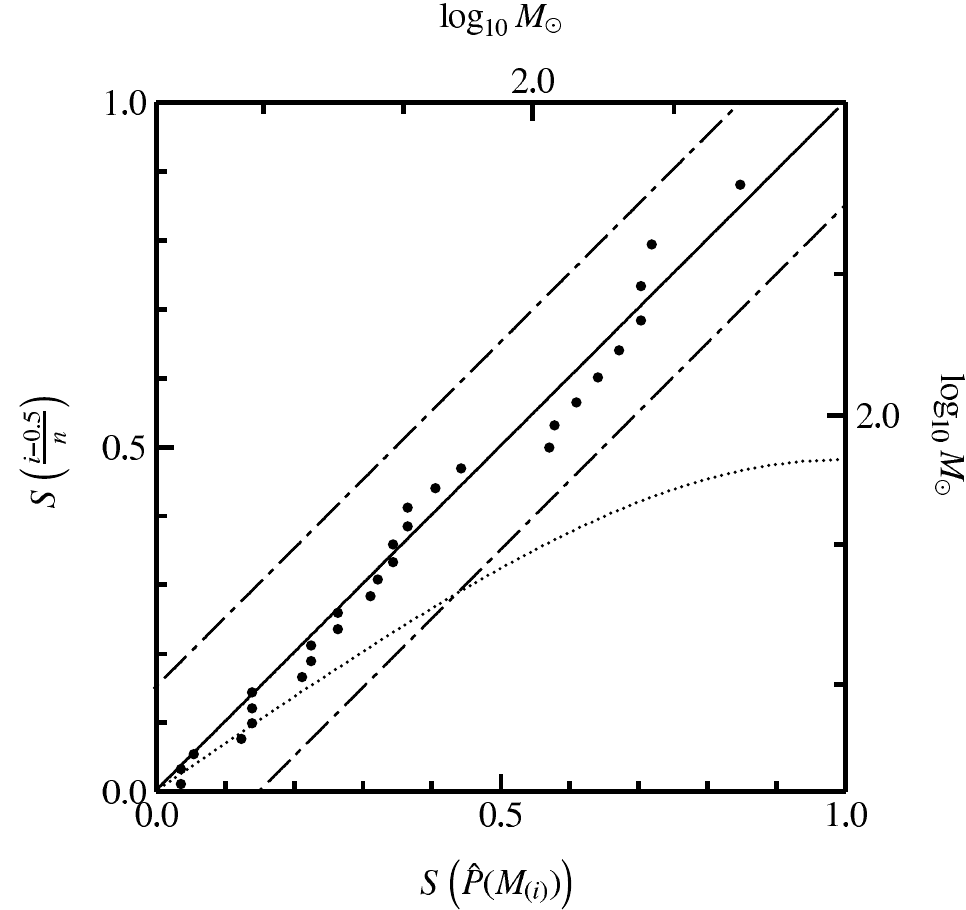}
\caption{\label{r136truncchlebowski}  Truncated SPP plot of the massive stars in R136 with masses according to the model of \citet{chlebowski+garmany1991} and parameters estimated using the modified ML method.
Also shown is the curve for a infinite power law (dotted).
The parallels to the diagonal limit the acceptance region of the stabilised KS test, null hypothesis of a truncated power law, significance level 5\%.}
\end{figure}

As a first exemplary application of the presented statistical techniques, in particular of the estimators, we chose the data set of massive stars in R136 published by \citet{massey+hunter1998}.
They gave for the 29 most massive stars the masses based on two different stellar models (\citealp{chlebowski+garmany1991}, with masses ranging from 56 $\Msun$ to 136 $\Msun$, and \citealp{vacca-etal1996}, $75-155\ \Msun$).
The results of the estimators are shown in Table \ref{r136results} where a Monte-Carlo sample of size 10000 was used to calculate the standard deviations.

Beg's estimator and the modified ML method agree well ($\hat{\alpha}=2.2$), the ML estimate is slightly smaller ($\hat{\alpha}=2.1$).
In reasonable agreement with this value are also the results of variable-size binning and fitting the complementary cumulative DF plot.
For comparison the results of \citet{koen2006} are also given in Table \ref{r136results}. 
The ML estimates are equal, only the CCDF result differs, likely due to a different definition of the empirical DF (Koen uses $i/(n+1)$ whereas here $(i-0.5)/n$ is used).
The ML method without including an upper limit  gives a much larger exponent ($\hat{\alpha}=3.5$) which shows the effect of a model mismatch.
A comparison of this value only with a constant-size histogram, where a linear regression gives $\hat{\alpha}=3.4$, would not give any indication of the mismatch. 

The upper limit is determined as $\approx 140\ \Msun$ by Beg's estimator, the ML and the CCDF method.
The results from variable-size binning are not consistent with the data set, because this upper limit is smaller than the largest data point.

For the goodness-of-fit analysis an SPP plot with a truncated power law as null hypothesis is shown in Fig. \ref{r136truncchlebowski}. 
The curve for the infinite power law is clearly not fitting the data.
The stabilised Kolmogorov-Smirnov, Cram\'er-von Mises and SPP correlation coefficient test all give a strong disagreement of the data with an infinite power law and no disagreement with a truncated power law.

The modified ML estimates from the data set using the models of \citealp{vacca-etal1996} are $\hat{\alpha}=2.87\pm0.98$ and $\hat{M}_\MAX=163\pm 9 \ \Msun$. 
The goodness of fit tests indicate a truncated power law with high significance too.

\subsection{The young star clusters in the Large Magellanic Cloud}
\begin{figure}
\includegraphics[width=8.5cm]{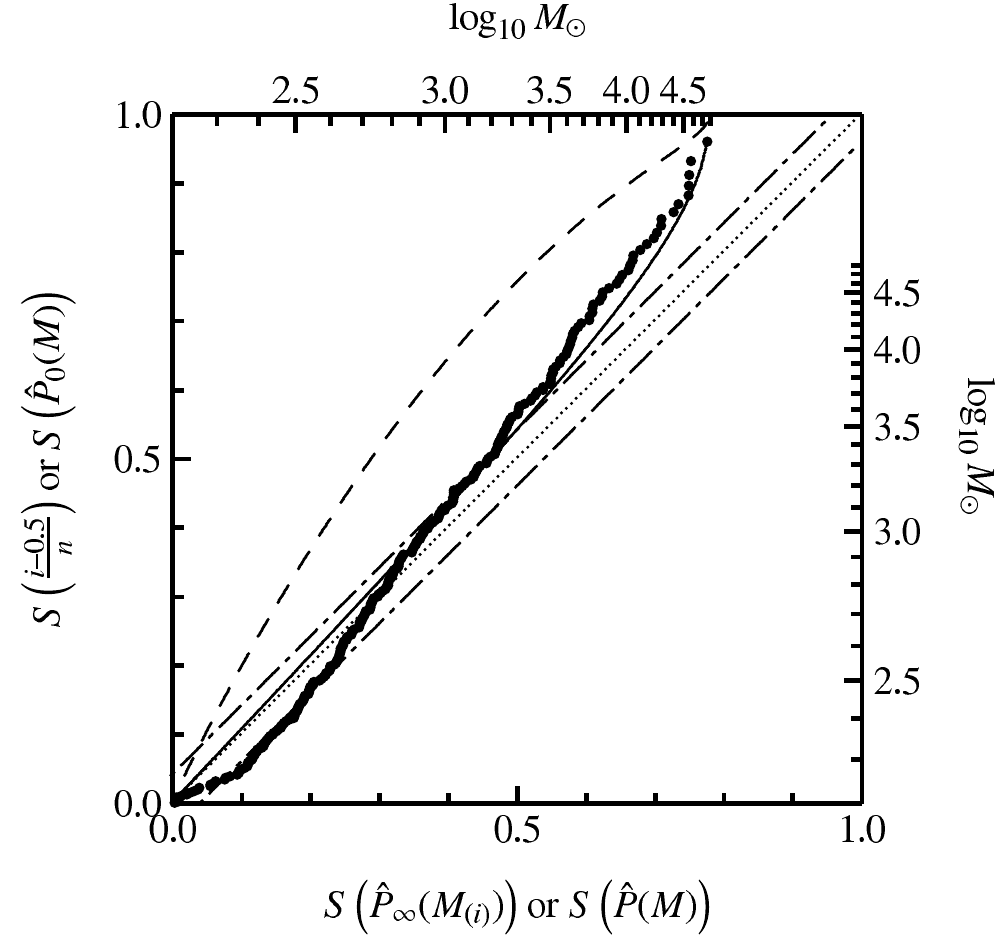}
\caption{\label{lmc-degrijsinf}  Infinite SPP plot of the LMC star clusters (age $< 10^{7.5}$ yr) with the lower mass limit of \citet{degrijs+anders2006}, $10^{2.2}\ \Msun$. (Dotted line: infinite hypothesis, $\hat{\alpha}_\MML=1.47$; solid curve: truncated hypothesis, parameters as estimated ($\hat{\alpha}=1.47$, $\hat{M}_{\MAX,\MML}=64200\ \Msun$
); dashed curve: truncated hypothesis, $\alpha=2$, $\hat{M}_{\MAX,\MML}=64200\ \Msun$; dash-dotted lines: limits of the acceptance region of the stabilised Kolmogorov-Smirnov test, significance level 5\%). }
\end{figure}

\begin{figure}
\includegraphics[width=8.5cm]{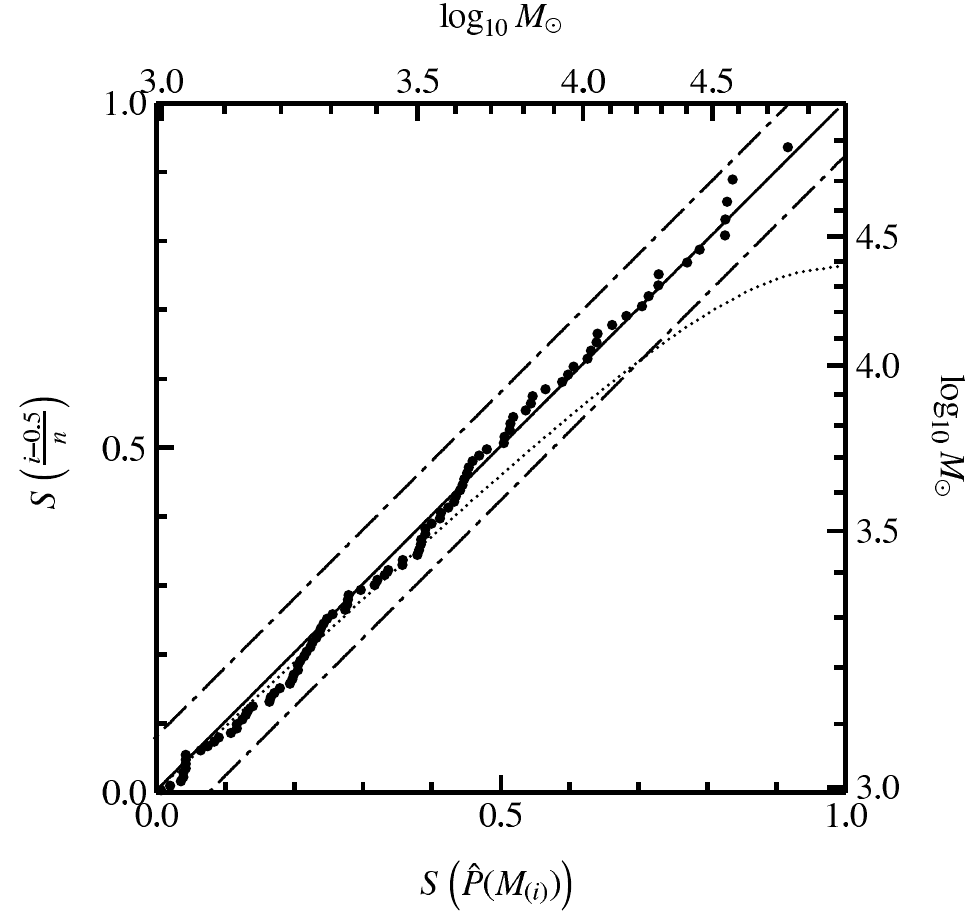}
\caption{\label{lmctrunc}Truncated SPP plot of the LMC star clusters (age $< 10^{7.5}$ yr) starting at $10^{2.5}\ \Msun$. (Dotted line: infinite hypothesis, $\hat{\alpha}_\MML=1.62$; solid line: truncated hypothesis, parameters as estimated ($\hat{\alpha}_\MML=1.62$, $\hat{M}_{\MAX,\MML}=68000\ \Msun$
); dashed: truncated hypothesis, $\alpha=2$, $\hat{M}_{\MAX,\MML}=68000\ \Msun$; dash-dotted: limits of the acceptance region of the stabilised Kolmogorov-Smirnov test, significance level 5\%). }
\end{figure}

\begin{figure}
\includegraphics[width=8.5cm]{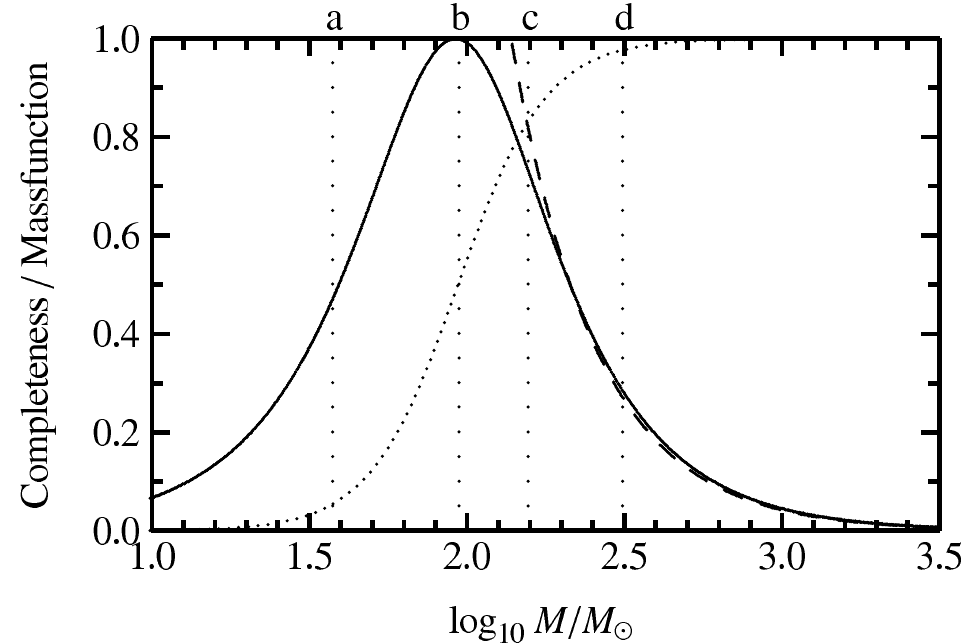}
\caption{\label{lmccompleteness} Influence of the completeness (dotted line) on the observable mass function (solid line), based on an assumed power law (dashed line) as the underlying distribution function. The parameters were chosen to match the situation for the Large Magellanic Cloud, see text. The vertical lines correspond to a: $10^{1.58}\ \Msun$, b: $10^{1.98}\ \Msun$, c: $10^{2.2}\ \Msun$, and d: $10^{2.5}\ \Msun$.
The mass function is scaled arbitrarily for better visibility.
}
\end{figure}

The second example for the methods presented above, with an emphasis on the advantage of the SPP plot, is the analysis of the mass distribution of young star clusters in the Large Magellanic Cloud.
We use a part of the data set given by \citet{degrijs+anders2006}, the star clusters with ages younger than $10^{7.5}$ yr and masses larger than $10^{2.2}\ \Msun$.

Based on an inspection of the shape of a histogram of the data \citet{degrijs+anders2006} concluded that there is a significant flattening of the mass function for $M <10^3\ \Msun$ (see their fig. 8).
Indeed, also an SPP plot with an infinite null hypothesis, Fig \ref{lmc-degrijsinf}, shows that the empirical curve of the data is strongly bent in the lower mass range ($M \lesssim 10^{2.5}\ \Msun$, estimated exponent $\approx 1.5$).
In the upper mass range the infinite SPP plot reveals that the data are better described by a truncated power law (solid line).
This indicates that above $\approx 10^{2.5}\ \Msun$ the data presumably will be consistent with a truncated power law.

With an SPP plot using only the star clusters more massive than $\approx 10^{2.5}\ \Msun$ this hypothesis is confirmed (Fig. \ref{lmctrunc}), the stabilised Kolmogorov-Smirnov acceptance region is not exceeded.
Thus, a change in the slope or shape of the mass function in the mass range $10^{2.5}-10^3\ \Msun$, as stated by \citet{degrijs+anders2006}, cannot be deduced using our techniques.
Only the mass range $10^{2.2}-10^{2.5}\ \Msun$ seems to deviate from the power law.
The slope which is derived from the data with masses larger than $10^{2.5}\ \Msun$ is $\hat{\alpha}=1.6\pm0.06$ and an upper limit $68\pm11 \times 10^3 \Msun$ is obtained by the modified Maximum Likelihood method.
This exponent is smaller than the value determined by \citet{degrijs+anders2006}, $\alpha=1.8\pm0.1$, who used constant-size binning and star clusters more massive than $10^3\ \Msun$ (for this mass range the modified maximum Likelihood estimate is $\hat{\alpha}=1.63\pm 0.1$).

The feature in the mass range $10^{2.2}-10^{2.5}\ \Msun$ could be caused by an actual change of the mass function.
However, since it is at the lower mass end, it could also be caused by an incomplete data set.
The completeness limit adopted by \citet{degrijs+anders2006} was derived by \citet{hunter-etal2003} from the behaviour of the luminosity function (see fig. 4 of \citealp{hunter-etal2003}).
They used as the brightness limit the brightness where the luminosity function reaches at the faint side half of its peak value, obtaining $\mathcal{M}_V = -3.5\ \mathrm{mag}$ or $10^{1.58}\ \Msun$ (using $M=10^{6+0.4(-14.55-\mathcal{M}_V)}\ \Msun$, \citealp{hunter-etal2003}, eq. 1).
The mass related to the brightness limit is valid for clusters of an age of 10 Myr.
In a similar way \citet{parmentier+degrijs2008} derive -- starting from the mass distribution of a chosen age interval older than 10 Myr -- from the mass which separates the lower 25\% from the upper 75\% a completeness limit of $\mathcal{M}_V = -4.7\ \mathrm{mag}$.
If we use this value also for younger clusters, then a completeness mass of $10^{2.06}\ \Msun$ would result (However, for unknown reasons an application of their method to clusters younger than 10 Myr leads to a different completeness mass of $10^{2.35}\ \Msun$, Parmentier, priv. comm.).
A completeness limit derived in such a way coincides approximately with the peak of the observed mass function.
But the transition from no detection to complete detection is smooth and has a certain broadness in which only a fraction of all sources is detected which can affect a wide mass range, illustrated in Fig. \ref{lmccompleteness}.
The observable mass function (solid line) is the product of the actual mass function (dashed, exponent $\alpha=1.6$) and the completeness function (dotted).
As the functional form for the completeness function we chose
\< c(M) &=& 1 - \left( 1 + \left( \frac{M}{M_0} \right)^\phi \right)^{-1}. \>
The parameters of the completeness function were chosen such that the half peak point of the observable mass function is at $10^{1.58}\ \Msun$ (or $\mathcal{M_V}=-3.5 \mathrm{mag}$, as \citealp{hunter-etal2003}, point a in Fig. \ref{lmccompleteness}) and the peak mass is $\approx 10^2\ \Msun$ ($\mathcal{M}_V \approx -4.5$, point b in Fig. \ref{lmccompleteness}). 
It is just a coincidence that for the used parameters ($\log_{10} M_0 =1.98$ and $\phi=3.12$) the 50\% completeness mass of the completeness function coincides with the peak mass of the observable mass function.
With these empirically determined parameters the observable mass function is shallower than the actual power law in the mass range below $\approx 10^{2.5}\ \Msun$.
This strongly supports the argument that the deviation of the data from the power law in Fig. \ref{lmc-degrijsinf} is caused by incompleteness.
The distribution of star clusters with ages $< 10^{7.5}$ yr are well consistent with a single power law with $\hat{\alpha}=1.6$, starting from $10^{2.5}\ \Msun$.

\section{Summary and conclusions}
In this work we compared methods to estimate the exponent and upper limit of a truncated power law distribution.
We reviewed graphical methods to represent the data.
Finally we studied goodness-of-fit tests, specifically to test for truncation.

Our results are:
\begin{enumerate}
\item A generally working estimator for the exponent and upper limit is our modified maximum likelihood method.
It performs well with respect to bias and standard deviation.

\item A maximum likelihood estimate of the exponent without considering a truncation can lead to biased results if the data stem from a truncated power law.

\item The estimator of \citet{beg1983} is also performing well but is numerically not stable. Variable-size binning as introduced by \citet{maizapellaniz+ubeda2005} performs well for the exponent.
The estimate for the upper limit based on the normalisation constant is biased.

\item The stabilising transformation introduced by \citet{michael1983} enhances plots and goodness-of-fit tests. For one-sided distributions only a half transformation should be made to achieve optimal results.

\item The stabilised PP plot is a particular useful display of the data.

\item The stabilised Kolmogorov-Smirnov statistic ($SD$), the stabilised PP correlation test ($Sk^2$), the stabilised Cram{\'e}r-von Mises statistic ($SC^2$), the Jackson statistic ($T$) and the QQ correlation ($r^2$) test are powerful goodness-of-fit tests for the truncated power law.

\item The exceedance statistic ($X$) is the most powerful test for truncation. Since it does not test for power-law behaviour it has to be used in combination with a powerful goodness-of-fit test for the truncated power law, as the ones mentioned in the previous point. 

\item The massive stars in R136 are well described by a truncated power law with $\hat{\alpha}=2.20\pm0.78$ and $\hat{M}_\MAX=143\pm9\ \Msun$, using the \citet{chlebowski+garmany1991} stellar models for mass determination, or $\hat{\alpha}=2.87\pm0.98$ and $\hat{M}_\MAX=163\pm9\ \Msun$, using the \citet{vacca-etal1996} stellar models. 

\item The young star clusters in the Large Magellanic Cloud (ages younger than $10^{7.5}$ yr) with masses larger than $10^{2.5}\ \Msun$ are well described by a truncated power law with $\hat{\alpha}=1.62\pm0.06$ and $\hat{M}_\MAX =68.8\pm11.6\ \times 10^3\ \Msun$.

\item A change in shape of the star cluster mass function in the Large Magellanic cloud in the low mass range $M < 10^3\ \Msun$, as reported by \citet{degrijs+anders2006}, cannot be verified. For $M>10^{2.5}\ \Msun$ the observed distribution follows a truncated power law, a flattening below $10^{2.5}\ \Msun$ is most likely caused by an underestimated completeness limit.

\end{enumerate}

\section{Acknowledgements}
We thank Cathie Clarke and Douglas Heggie for critical reading of the manuscript and valuable comments.
TM acknowledges financial support by the AIfA.

\bibliographystyle{mn2e}
\bibliography{stat,cluster}

\appendix

\section{Beg's estimator}

The estimator for the exponent \citep{beg1983} is given in its original form as
\< \hat{\theta} &=& \frac{
(n-3)! \sum_{j=0}^{j^\star} (-1)^j {n-2 \choose j} K_{j+1}^{n-4}
}{
(n-4)! \sum_{j=0}^{j^\star} (-1)^j {n-2 \choose j} K_{j+1}^{n-3}
} \label{thetabeg}, \>
where $\hat{\theta}=\hat{\alpha}-1$ and $K_j= T - nY - j(Z-Y)$ with $Y=\log_e X_{(1)}$, $Z=\log_e X_{(n)}$ and $T=\sum_{i=1}^n \log_e X_j$.
The terminating index of the sum, $j^\star$, is determined by the condition $T -nY - j(Z-Y) > 0$, as shown by \citet{beg1983}.
The estimate for the exponent is then $\hat{\alpha} = \hat{\theta} + 1$.

The direct evaluation of eq. \ref{thetabeg} involves the calculation of ${n-2 \choose j}$, which is only practicable for less than about 170 data points in double precision arithmetic.
This problem can be handled with a recursive implementation of the estimator, feasible for any number of data, as follows.

To abbreviate we introduce   $L_j = \left( 1-j  \frac{Z-Y}{T-nY} \right)$ which leads to
\< K_j^{n-4} &=& (T- nY)^{n-4} L_j^{n-4}. \>
With changing the limits of the sum and omitting $(n-3)!$ the numerator of eq. \ref{thetabeg} reads
\< - (T- nY)^{n-4} \sum_{j=1}^{j^\star} (-1)^j {n-2 \choose j-1} L_j^{n-4} .\>
Omitting the prefactor $(T-nY)^{n-4}$ , the expanded sum reads
\< - \underbrace{\frac{(n \!\! - \!\! 2)!}{0! (n \!\! - \!\! 2)!}}_{=1} L_1^{n-4} 
\! + \! \frac{(n \!\! - \!\! 2)}{1 } L_2^{n-4} 
\! - \! \frac{(n \!\! - \!\! 2)(n \!\! - \!\! 3)}{1\cdot2} L_3^{n-4} 
\!+ \dots .\>
Starting with the second term this can be written as
\< \underbrace{\frac{n-2}{1} \left( L_2^{n-4} - 
\!\!\!\! \underbrace{ \frac{n-3}{2}\left( L_3^{n-4} - 
 - \dots 
 \right)}_{=:S_3^{(n-4)}}
  \right)}_{=:S_2^{(n-4)}  } .\>
The superscript $(n-4)$ should only indicate the exponent and is not used as an exponent in $S_j^{(n-4)}$.
From this the recursion can easily be seen:
\< S_{j'-1}^{(n-4)} &=& \frac{n-j'}{j'-1} \left( L_{j'}^{n-4} -  S_{j'}^{(n-4)}
\right), \label{recn-4} \>
where $j'$  descends from $j^\star$ to 2 and $S_{j^\star}=0$.
The last step is 
\< S^{(n-4)} := S_1^{(n-4)} = L_1^{n-4} - S_2^{(n-4)}. \label{lastn-4} \> 
The recursion for the denominator in eq. \ref{thetabeg} is as for the numerator, but replacing the exponent $n-4$ by $n-3$ in equations \ref{recn-4} and \ref{lastn-4}.
The estimator of $\theta$ is then (remembering all omitted factors)
\< \hat{\theta} &=& \frac{n-3  }{T-nY} \frac{S^{(n-4)}}{S^{(n-3)} }.\>

The estimators for the upper limit in the form of \citet{beg1983} is 
\< \hat{x}_\MAX \!\!\! &=& \!\!\! X_{(n)} \left( 1 + \frac{1}{n(n-1)} \frac{ 
\sum_{j=0}^{j^\star} (-1)^j { n-1 \choose j} K_j^{n-2} 
}{
\sum_{j=0}^{j^\star} (-1)^j { n-2 \choose j} K_{j+1}^{n-3} 
} \right). \>
The recursion formula for the sum in the numerator follows by analogous steps as before with
\< S_j^{'(n-2)} &=& \frac{n-j}{j} \left( L_j^{n-2} - S_j^{'(n-2)} \right), \>
 the last step
\<  S^{'(n-2)} &=& 1 - S_1^{'(n-2)}.\>
and
\< \hat{x}_\MAX &=& X_{(n)} \left( 1 + \frac{(T - nY)}{n(n-1)} \frac{ S^{'(n-2)} }{ S^{(n-3)}} \right), \>
with $S^{(n-3)}$ from the estimator for the exponent above.

\label{lastpage}
\bsp

\end{document}

%% file: power.tex
 \begin{tabular}{r|r|
 *{ 32 }{r}}
 n & $\beta$ & 
 $D$&
 $SD$&
 $C^2$ &
 $SC^2$ &
 $A^2$ &
 $r^2$ &
 $k^2$ &
 $k_0^2$ &
 $Sk^2$ &
 $Sk_0^2$ &
 $W$&
 $T$&
 $\Lambda$&
 $X$
 \\\hline
 \hline
 \multicolumn{5}{l}{
 10.
 150.
 }\\
 33 &
       1.7&
      54.1&
      72.0&
      59.9&
      69.4&
      60.4&
      51.7&
       7.7&
      59.9&
      17.4&
      65.9&
      47.4&
      56.1&
      64.6&
     100.0&
 \\
 50 &
       1.7&
      70.9&
      88.3&
      75.0&
      86.0&
      78.7&
      69.9&
       5.1&
      75.3&
      16.3&
      87.3&
      68.7&
      78.6&
      86.3&
     100.0&
 \\
 99 &
       1.7&
      93.8&
     100.0&
      96.8&
      99.8&
      98.6&
      98.2&
       3.7&
      96.3&
      40.3&
      99.9&
      96.7&
      98.6&
      99.9&
     100.0&
 \\
 33 &
       2.0&
      23.3&
      29.6&
      24.4&
      28.2&
      26.0&
      22.9&
       4.5&
      26.2&
       7.0&
      28.2&
      22.2&
      23.9&
      29.8&
      53.4&
 \\
 50 &
       2.0&
      25.0&
      58.2&
      32.8&
      46.3&
      36.2&
      31.3&
       3.7&
      32.4&
       7.8&
      50.4&
      38.8&
      43.5&
      61.0&
     100.0&
 \\
 99 &
       2.0&
      39.1&
      93.7&
      50.7&
      80.4&
      63.7&
      58.7&
       3.1&
      50.3&
      11.5&
      85.2&
      68.9&
      74.8&
      94.9&
     100.0&
 \\
 33 &
       2.3&
       7.8&
      11.5&
       8.9&
      10.5&
       9.2&
       5.9&
       5.4&
       9.7&
       5.5&
      10.5&
      10.7&
      11.7&
      11.3&
      11.4&
 \\
 50 &
       2.3&
       8.5&
      18.3&
       8.8&
      14.1&
       9.3&
       4.0&
       2.9&
       9.2&
       3.7&
      15.0&
      14.2&
      15.0&
      18.9&
      22.3&
 \\
 99 &
       2.3&
      11.9&
      48.9&
      11.5&
      30.4&
      15.6&
       4.8&
       4.2&
      14.1&
       4.9&
      31.1&
      29.3&
      33.4&
      53.5&
      80.4&
 \\
 \hline
 \multicolumn{5}{l}{
 10000.
 1000000.
 }\\
 33 &
       1.7&
      14.3&
      18.5&
      15.3&
      19.1&
      14.5&
      18.8&
       6.6&
      15.3&
       7.5&
      19.3&
      20.0&
      17.9&
      17.6&
      19.5&
 \\
 50 &
       1.7&
      12.7&
      33.3&
      14.9&
      25.6&
      17.8&
      28.3&
       5.2&
      18.4&
       6.5&
      26.5&
      23.8&
      25.6&
      31.3&
      50.9&
 \\
 99 &
       1.7&
      19.9&
      77.4&
      29.1&
      57.3&
      40.1&
      60.1&
       4.2&
      30.9&
       8.0&
      61.8&
      52.1&
      58.0&
      76.9&
     100.0&
 \\
 33 &
       2.0&
       5.3&
       5.7&
       4.8&
       5.6&
       5.8&
       5.9&
       6.5&
       4.8&
       5.1&
       5.9&
       7.0&
       6.8&
       5.4&
       6.2&
 \\
 50 &
       2.0&
       6.8&
       8.4&
       7.6&
       7.8&
       7.2&
       5.5&
       5.7&
       8.3&
       4.1&
       7.3&
       8.2&
       8.8&
       8.6&
       8.7&
 \\
 99 &
       2.0&
       3.4&
      14.5&
       3.4&
       6.2&
       3.9&
       5.3&
       3.9&
       3.7&
       5.0&
       9.8&
      12.0&
      11.7&
      16.6&
      12.6&
 \\
 33 &
       2.3&
       5.4&
       4.9&
       4.9&
       4.4&
       4.5&
       3.3&
       4.7&
       4.1&
       4.5&
       4.2&
       4.4&
       4.6&
       4.9&
       4.9&
 \\
 50 &
       2.3&
       4.2&
       6.0&
       5.2&
       5.9&
       5.3&
       2.2&
       3.4&
       4.4&
       3.9&
       5.5&
       4.6&
       4.9&
       6.5&
       5.5&
 \\
 99 &
       2.3&
       5.8&
       7.0&
       6.8&
       6.9&
       7.2&
       1.1&
       6.5&
       6.0&
       6.0&
       6.9&
       8.6&
       8.6&
       7.8&
       6.8&
 \\
 \end{tabular}